\documentclass{article}
\usepackage{graphicx}
\usepackage{seqsplit}
\usepackage{enumerate}
\usepackage{hyperref}
\usepackage{xcolor}
\usepackage{pagecolor}
\definecolor{backgroundcolor}{HTML}{FFFAE6}
\pagecolor{backgroundcolor}
\title{Gateways for Institutional-Grade Commerce and Interoperability of Digital Assets}
\author{Rafael Belchior, Thomas Hardjono, Alex Chiriac, \\ Venkatraman Ranakrishna}

\date{January 2025}

\begin{document}

\maketitle
\noindent This document is a technical high-level version of the Secure Asset Transfer Protocol \cite{Hargreaves_Hardjono_Belchior_2023} whitepaper.

\section{Executive Summary}

It is time for the legacy financial infrastructure to seamlessly connect with modern, decentralized infrastructure \cite{Hardjono_Lipton_Pentland_2019}. Although it is increasingly evident that decentralized infrastructure for finance (namely distributed ledgers) will coexist with and complement legacy infrastructure, it is also clear that such interoperability efforts carry new risks and concerns \cite{SoKSecurityPrivacy,Belchior_Vasconcelos_Guerreiro_Correia_2021}. 
In particular, managing the range of heterogeneous (and not well-established) infrastructure brings security, privacy, and regulatory issues. 
The first step to overcome some of these challenges is to recognize that in many deployment instances using distributed ledgers,
the purpose of the ledger is to share resources among the community members. 
In other words, there needs to be recognition that {\em borders} exist
between communities and, therefore, between the distributed ledgers
utilized by each community.
These borders could also be present due to regulatory requirements within
some national jurisdictions.
The second step after recognizing that borders exist is to understand
that interoperability across systems can be best achieved 
through the use of {\em standardized service interfaces} 
(or application programming interfaces (API))~\cite{HardjonoLipton2024-Service-API}.
Service APIs are a well-known construct within Web2 systems
that enable better security and improved business cost savings
by providing stable interfaces into systems and networks.
External entities are assured that the services
obtained via the standardized APIs will deliver
the predictable outcome,
without these entities needing to know the technical implementation details
of the systems behind the APIs.

In this paper we use the term \emph{ledger gateways} (or simply {\em gateways}) 
to denote the computer and software systems that implement 
the standardized service interfaces
into a distributed ledger.
The main purpose of a gateway is to communicate with other {\em peer} gateways
that implement the same standardized service interface.
Among others, peer gateways perform the transfer of data and value
across borders (legal or national borders).
Gateways also become a mechanism to manage
a permissioned environment, 
where abiding by laws and regulations is crucial for business compliance
(e.g., EU General Data Protection Regulations (GDPR) \cite{eu2016gdpr}, EU MiCa regulation on digital assets \cite{eu2020mica, ESMA}, FAFT Recommendation 15 \cite{fatf2018recommendation15}, ISO 27001 \cite{iso27001}).

Thus, our proposal aims to solve the interoperability problem
facing many digital asset networks that utilize distributed ledgers using gateways and APIs as interconnection points.
One major challenge today with some distributed ledgers
is that assets (i.e. tokenized value)
in one network cannot be moved easily to another network
because of the need for some atomic actions to be carried out
across these independent asset networks.
The problem is more acute in the case of private asset networks, 
where external entities have no visibility into
the state of an asset in the private network. 
An example is regulated tokenized real-world private assets,
such as property ownership certificates and 
regulated central bank digital currencies (CBDC). 
A standard APIs could be defined that implements
a query-response protocol enabling
authorized external parties to query the status
of the asset within a private asset network.
The gateway would deploy this API,
providing external parties with a common way
to query the private asset network.
Another example
is the a need to transfer regulated tokenized assets
from one private asset network to another private asset network.
Work is underway in the Internet Engineering Task Force (IETF)
to standardize a common 
{\em Secure Asset Transfer Protocol} (SATP)~\cite{Hargreaves_Hardjono_Belchior_2023}
that operates through standardized APIs deployed at the gateways.
The aim of the SATP asset transfer protocol 
is to enable guaranteed unidirectional transfer between asset networks
while being agnostic (oblivious) to the {\em type} of the asset being transferred.

For institutional-grade interoperability between asset networks,
we envisage that several foundational components or parts will be needed:
\begin{itemize}

\item A clear definition of the distributed ledger gateway architecture, capabilities, and functionality; 

\item Support for standardized APIs, standard data (metadata) formats 
and standard asset transfer protocols (e.g. SATP);

\item Abstractions and tooling for implementing business 
logic using gateway-supported interoperability protocols.
\end{itemize}

\section{Introduction}

The blockchain industry is evolving at a quick pace, but more services need to be built on top of its technical foundations in order to deliver value to its users. 
The large number of Layer~1 and Layer~2 solutions available today
indicates a certain degree of commoditization of the functions provided by blockchain technology.
The Gartner Hype Curve from July 2024~\cite{gartner2024}
places the blockchain industry at the stage of the ``trough of disillusionment",
which means, among others,
that further technical developments and standardization
will be needed if blockchain technology is to climb the incline of productivity.
Out of the current proliferation of the different blockchains -- many experimenting with their
own decentralization technical approaches --
those that prioritize high-throughput and low-latency
transactions~\cite{Belchior_Sußenguth_Feng_Hardjono_Vasconcelos_Correia_2023}
may emerge to be the dominant solution.
These will be the
``facilitators'' of the new emerging tokenization trend (e.g., tokenization of physical assets), 
which is predicted to reach tens of trillions of US dollars in the near future~\cite{coindesk2024tokenization}.

One major obstacle today with regard to the broad adoption of blockchain technology 
is the absence of {\em interoperability} across these blockchains
from the perspective of the tokenized value.
Tokens that represent value or rights -- as understood by the EU MiCA regulation~\cite{eu2020mica, ESMA} --
must have the ability to be transferred across blockchain,
thereby preventing the tokenized value and the economic actors who utilize these tokens
from being captured by a specific blockchain.

Another emerging trend is the establishment of {\em private} blockchain networks
by trading communities or consortiums that wish to reduce operational costs
and provide shared transparency into the state of the digital assets relevant 
to each community~\cite{JPMCoinSystem}.
Thus, the challenge of interoperability is further compounded by the fact that
some of these blockchain networks will be private,
meaning that their ledger is closed and not visible to non-members and external entities.
On the technical plane, this means that solutions that function well
within the public/open blockchains may not be useful 
for interoperability with private blockchain networks.

A third challenge pertains to the {\em compatibility} of the tokenized value or rights
across different legal jurisdictions and economic zones.
At the heart of this challenge is the need for a standard codification of the definition of assets that
are tokenizable~\cite{eu2020mica}.
The asset definitions must be expressed using a common standardized data {\em schema}
which must be written in a machine-readable format (e.g., JSON).
This data schema for a given asset class allows computers and nodes around the world
to reach a shared semantic understanding of the nature of the asset being tokenized.

In order to solve this ``tri-lemma" of problems and many others,
a correct, robust, and scalable architecture for the interoperability of blockchain networks
must be developed that enable the numerous competing and cooperating blockchains
to transfer tokenized valuer across these blockchain networks in a seamless fashion.
The TCP/IP Internet is able today to handle  billions of connections each day
across the globe by virtue of its scalable architecture.
As such, it is wise for the blockchain industry to re-use
many of the engineering constructs present in the Internet Architecture
to solve the interoperability problem~\cite{Hardjono_Lipton_Pentland_2019}.

One such construct is the {\em cross-domain gateways}
which enables local networks to interoperate with one another
while retaining their independence in running their internal protocols.
In the context of tokenized assets and blockchain networks,
the gateway model based on the Internet Architecture 
bounds each blockchain network 
by special nodes called {\em transfer gateways} (or simply ``gateways'')
that orchestrate and coordinate cross-network transfers of data/metadata and tokenized assets. 
Two gateways in different networks {\em peer} with one another
when transferring a tokenized asset from one to the other.
The peer gateways run a special protocol between them
called the {\em Secure Asset Transfer Protocol} (SATP)~\cite{Hargreaves_Hardjono_Belchior_2023}.
Each gateway represents one asset network\footnote{A gateway can be decoupled from a network and can represent different networks at different times. For the sake of simplicity, we present the one-network case.}, 
and the SATP protocol enables gateways to perform a transfer of tokenized assets
from the origin blockchain network to a destination network in a safe and consistent manner.
For any given transfer,
the evidence of the transfer can be verified by a third-party audit in the case of disputes.

Since gateway nodes (which are computers) may be unavailable for one reason or another (e.g., crash),
each blockchain network must deploy several gateway nodes for higher cross-network throughput.
Should a transfer instance between two peer gateways become interrupted due to the sudden
unavailability of one of the gateways,
a crash-recovery protocol has also been devised~\cite{Belchior_Correia_Augusto_Hardjono_2024} 
to accompany the SATP protocol.
In the event of a crash, the gateways can safely recover and continue the execution of the SATP protocol.
If the crash is severe and the session between the peers cannot be recovered or resumed,
a rollback of the transfer session is conducted
in such away that both the origin network and destination network remain in a consistent state.

A key requirement for transferring assets from one blockchain network to another
is to ensure that the token (representing the tokenized asset)
is valid in one network only at any given time. 
This means that the SATP protocol must ensure that the properties of 
atomicity, consistency, isolation, and durability (ACID) of the underlying networks\footnote{Technically, these properties are also enforced globally, i.e., across independent networks} are satisfied in an asset transfer. Commitments and rollbacks must be supported in the case of an asset mid-transfer failure \cite{Belchior_Vasconcelos_Correia_Hardjono_2022}.

\section{Gateway Architecture}

Though the blockchain gateway infrastructure can run a variety of protocols, we focus on SATP in this paper as it is the most mature effort toward standardization of such protocols. In turn, there is an applicational layer on top of SATP. Our architecture relies on different sets of APIs (quoting \cite{hardjono2023standardized}):

\begin{figure*}[h]
    \centering
    \includegraphics[scale=0.38]{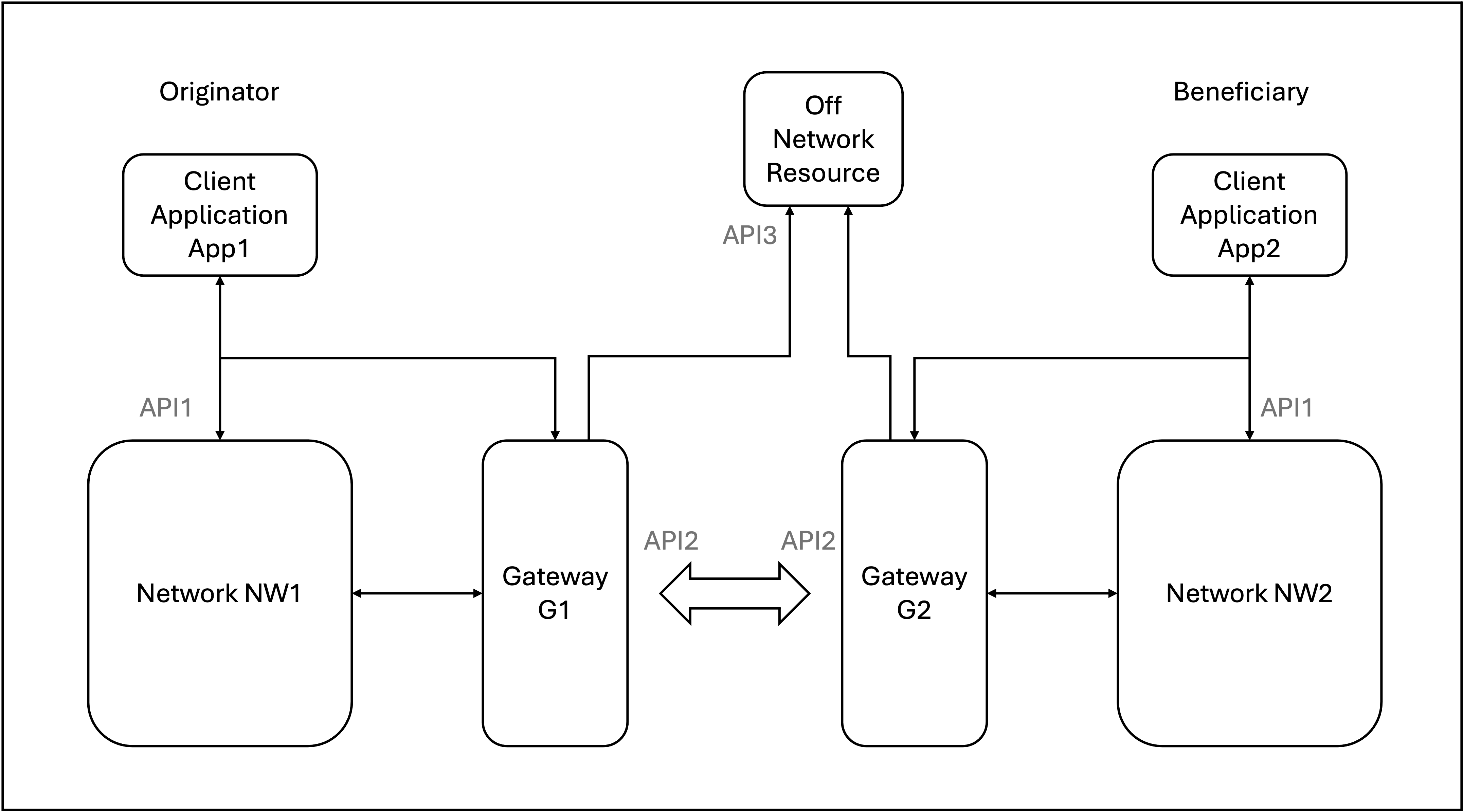}
 \caption{Summary of the API model for tokenized asset networks.}
    \label{fig:2}
\end{figure*}

\begin{enumerate}

\item Application-to-Network APIs: These are interfaces an asset network (i.e., blockchain-based) must expose to applications seeking to perform tasks (e.g., asset transfers from Alice to Bob). Included here are APIs that address pre-transaction validation of the various aspects of the proposed transaction before processing and transfer. In other words, these are the APIs for “pre-flight” checks. These are shown as API1 in Figure \ref{fig:2}.

\item SATP and Network Interoperability APIs: This is the set of APIs used for connecting blockchains via gateway nodes. This includes APIs for checking network-level information (e.g., blockchain identifiers, state proofs) as well as APIs for cross-network transfers of assets. These are shown as API2 in Figure \ref{fig:2}.

\item Common APIs for off-chain asset-related artifact validation: These are the APIs on the side of the off-chain service that enable the validation of certain aspects of a proposed transaction. For example, in the case of a proposed transfer of an asset from Alice to Bob, the identities of Alice and Bob must first be verified before the transaction is proposed to the network. The identity provider (IdP) industry has already specified and deployed standardized service interfaces to perform these checks. These are represented in as API3 in Figure \ref{fig:2}.

\end{enumerate}

\section{Gateway Service Providers}

In the following sections, we will describe key points that a Gateway Service Provider (GSP) could satisfy. The recommendations on this subject are based on our previous work from FATF Recommendation 15. \cite{hardjono2023standardized} First, let us recall the definition of a virtual asset from this recommendation: a digital representation of value that can be digitally traded. Building on this, a Virtual Asset Service Provider (VASP) was defined as a business that conducts one or more of the following activities (or operations for or on behalf of another natural or legal person or business):
\begin{itemize}
\item exchange of virtual assets for fiat currencies
\item exchange of different forms of virtual assets for each other
\item transfer of virtual assets
\item safekeeping and/or administration of virtual assets or instruments enabling control over virtual assets
\item participation in and provision of financial services related to an issuer’s offer and/or sale of a virtual asset
\end{itemize}

The implication of FATF Recommendation 15, among others, is that VASPs must be able to share the originator and beneficiary information for virtual asset transactions. This process – also known as the Travel Rule – originates from the US Bank Secrecy Act (BSA - 31 USC 5311 - 5330)\footnote{https://www.occ.treas.gov/topics/supervision-and-examination/bsa/index-bsa.html}, which mandates that each financial institution delivers certain types of information to the next financial institution when a funds transmittal event involves more than one such institution in a sequence. This customer information includes the following:
\begin{itemize}
\item originator’s name
\item originator’s account number (e.g., at the Originator’s VASP)
\item originator’s geographical address, national identity number, or customer identification number (or date and place of birth)
\item beneficiary’s name
\item beneficiary account number (e.g., at the Beneficiary-VASP)
\end {itemize}

\section{Operational Model for Gateway Service Provider} 

In the previous section, we read that VASPs can have legal legitimacy to operate distributed ledger gateways.  We present a deployment model where a Gateway Service Provider (GSP) owns and operates a set of gateways connected to one or more networks. The GSP implements a set of standardized APIs, allowing it to peer with other gateways that implement the same standard APIs for protocol execution, identification, asset profile retrieving, payment settlements, and blockchain connectivity, among others. Customers with existing IT systems who may not wish to implement their enterprise gateway may choose to obtain interoperability services via a GSP. This is similar to how the traditional Internet Service Provider (ISP) achieves connectivity to other ISPs on the Internet.

\begin{itemize}
\item The GSP can perform various services (such as identity verification, network type verification, and asset type checking) not usually performed by traditional financial institutions. Most banks may be prohibited from connecting directly to a public permission-less token network.
\item GSP can serve multiple private networks simultaneously, leveraging existing KYC/AML systems to check identities and asset token types. The GSP can provide broad connectivity to the growing asset token networks worldwide.
\item Small to medium-sized financial institutions can outsource to (i.e., buy services from) one or more GSPs instead of implementing SATP in-house, thereby reducing IT costs. 
\item The GSP can assist its customers (financial institutions) in fulfilling their annual legal reporting by providing verifiable histories of transactions related to those customers.

\end{itemize}

\section{Implementation and Evaluation}
Two open-source versions of the protocol are available under Hyperledger Cacti, the flagship interoperability project for distributed ledger networks\footnote{Typescript implementation available at \url{https://github.com/hyperledger/cacti/tree/main/packages/cactus-plugin-odap-hermes}; Rust implementation available here: \url{https://github.com/hyperledger/cacti/tree/main/weaver}}. We refer to the project source and to academic studies for technical details. Several closed-source implementations are being actively developed by IETF'S SATP group.

\section{Institutional Adoption and Use Cases}
Several well-recognized and established universities, research institutions, and blockchain companies are working on standardizing and implementing SATP to streamline next-generation payment systems and seamless interoperability. These standardization and engineering efforts have gathered attention from the industry and academia. The national Portuguese blockchain project \cite{BlockchainPT} has chosen SATP to provide interoperability across the project verticals. We have compiled a detailed list of use cases \cite{Ramakrishna_Hardjono_2023}.

\section{Open Challenges and Recommendations}
The blockchain interoperability research area identifies several critical unsolved challenges: 

\begin{itemize}
    
\item Privacy-preserving technologies: Existing literature suggests that unlinkability of transactions is possible only in the scenarios where privacy-preserving blockchains are involved (e.g., Hyperledger Fabric, ZCash, Monero, centralized systems) \cite{SoKSecurityPrivacy}. This would provide some confidentiality to the end-user. However, data privacy for non-private blockchains currently does not exist and is a research topic. Gateways could alleviate this problem and, at the same time, provide an acceptable level of decentralization.

\item Regulatory and compliance requirements: The status quo requires virtual asset providers to abide by several data privacy and asset management regulations. Such regulations impact global organizations that utilize centralized and decentralized technology to provide several services in the financial sector. There is a need for gateways to manage such complexity by translating requests from the application layer into legal-binding transactions across centralized and decentralized infrastructure.

\end{itemize}

\subsection*{Acknowledgments}
\small We thank all the contributors for the Secure Asset Transfer Protocol working group at IETF. 

\bibliographystyle{acm}

\bibliography{main}

\end{document}